\documentclass{PoS}

\usepackage{amssymb}
\usepackage{fontenc}
\usepackage{times}
\usepackage{mathptmx}
\usepackage{graphicx}
\usepackage{booktabs}
\usepackage{bm}
\usepackage{mathtools}
\usepackage{caption}
\usepackage{subcaption}
\usepackage{xfrac}
\usepackage{xcolor}

\let\OLDthebibliography\thebibliography
\renewcommand\thebibliography[1]{
  \OLDthebibliography{#1}
  \setlength{\parskip}{-1.4pt}
  \setlength{\itemsep}{-1.1pt plus 0ex}
}

\AtBeginDocument{\setlength\abovedisplayskip{1.4ex}}
\AtBeginDocument{\setlength\belowdisplayskip{1.4ex}}
\AtBeginDocument{\setlength\abovedisplayshortskip{1.4ex}}
\AtBeginDocument{\setlength\belowdisplayshortskip{1.4ex}}
\title{Matching of $N_f=2+1$ CLS ensembles to a tmQCD valence sector}

\ShortTitle{Matching of $N_f=2+1$ CLS ensembles to a tmQCD valence sector}

\author{
\begin{minipage}[b]{0.4\linewidth}
\includegraphics[height=2.5\baselineskip]{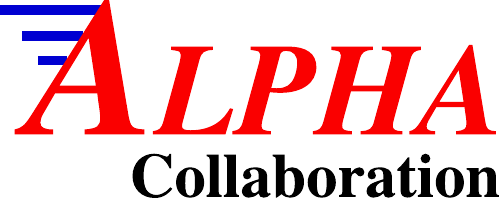}
\end{minipage}
\hfill
\begin{minipage}[b]{0.3\linewidth}
{\it
IFT-UAM/CSIC-19-12 \\
FTUAM-19-2}
\end{minipage}
}

\author{A.~Bussone$\,^{a,b}$, 
G.~Herdo\'{i}za$\,^{a,b}$, 
C.~Pena$\,^{a,b}$,
D.~Preti$\,^{c}$,
\speaker{J.\'{A}.~Romero}$\,^{b}$, 
J.~Ugarrio$\,\,^{a,b}$\\
\llap{$^a$}{Department of Theoretical Physics, Universidad Aut\'{o}noma de Madrid, E-28049 Madrid, Spain}\\
\llap{$^b$}{Instituto de F\'{i}sica Te\'{o}rica UAM-CSIC, c/ Nicol\'{a}s Cabrera 13-15, Universidad Aut\'{o}noma de Madrid, E-28049 Madrid, Spain}\\
\llap{$^c$}{INFN, Sezione di Torino
Via Pietro Giuria 1, I-10125 Turin, Italy}\\
E-mail:\,\email{ja.romero@csic.es}
}

\abstract{A mixed action composed of valence quark flavours regularized
with a fully-twisted tmQCD action and of $N_f=2+1$ flavours of
non-perturbatively ${\rm O}(a)$-improved Wilson sea quarks is described. Two
procedures for the matching of sea and valence quark masses are
discussed. We report about a comparison of the continuum-limit scaling
of pseudoscalar meson observables and of quark masses using the sea
and valence actions.}

\FullConference{The 36th Annual International Symposium on Lattice Field Theory - LATTICE2018\\
22-28 July, 2018\\
Michigan State University, East Lansing, Michigan, USA.}

\begin{document}

\section{Introduction}	\label{intro}
Lattice QCD computations in the heavy-quark sector are essential for the search of New
Physics. Precise determinations of decay constants and form factors involved
in charmed meson decays are fundamental to monitor the consistency
between the Standard Model predictions and upcoming experimental
results.
In particular, an accurate computation of D-meson semileptonic decay form factors is required to match the level of precision reached by experiments.
We consider a setup~\cite{Herdoiza:2017} aimed at addressing the leading systematic uncertainties in charm-quark observables.
In the sea sector, we employ a L\"{u}scher-Weisz tree-level improved gauge action and an $N_f=2+1$ Wilson Dirac fermionic action~\cite{Wilson:1974sk}, including the Sheikholeslami-Wohlert term~\cite{Sheikholeslami:1985ij}.
The gauge ensembles considered in this work, see Table~\ref{tab_ens}, were
produced within the CLS initiative~\cite{Bruno:2014jqa}. They lie along a line of constant trace of bare quark mass matrix,
\begin{equation}
  \mathrm{tr} M_\mathrm{q}=2m_{q,\ell}+m_{q,\mathrm{s}} = \mathrm{const}\, ,
  \label{tr_mb}
\end{equation}
where $m_{\mathrm{q,f}} = m_{0,\mathrm{f}}-m_\mathrm{cr}$. In this way, cut-off effects proportional to $M_q$ are constant for any quark mass.
In practice, it is beneficial to depart from the relation in
eq.~(\ref{tr_mb}) involving bare quark masses towards a renomalized chiral
trajectory in terms of following dimensionless quantities,
\begin{equation}
  \phi_2=8t_0\,m_\pi^2\,, \qquad \qquad  \phi_4=8t_0\left( m_K^2 + \frac{1}{2}m_\pi^2 \right) \,,
  \label{phi}
\end{equation}
in units of the gradient flow scale $t_0$.
This mass shift in the $u,d$ and  $s$ quark masses can be obtained through a low order Taylor expansion~\cite{Bruno:2016plf}.
Besides the ensembles laying over a chiral trajectory that approaches the physical point,
additional ensembles along a symmetric line with degenerate u, d and s
quark masses were generated at $\beta$=3.46 (see Table~\ref{tab_ens}).
Adopting open boundary conditions in the time direction allows to improve the sampling of configuration space at fine values of the lattice spacing~\cite{Luscher:2010iy,Luscher:2011kk,Luscher:2012av}.
\begin{table}[!htbp]
  \begin{center}
    \small
    \begin{tabular}{ccccccccc}
      \toprule
      Id &   $\beta$ & $~a$[fm] & $N_\mathrm{s}$  &  $N_\mathrm{t}$  & $m_\pi$[MeV] &   $m_K$[MeV] &  $m_\pi L$\\
      \midrule
      H101 & 3.40 & 0.087 & 32 & 96	& 420 &420  & 5.8\\
      H102 & 3.40 & 0.087 & 32 & 96 & 350 & 440 & 4.9\\
      H105 & 3.40 & 0.087 & 32 & 96	& 280 &460  & 3.9\\
      \midrule
      H400 & 3.46 & 0.077 & 32 & 96   & 420 & 420 & 5.2\\
      H401 & 3.46 & 0.077 & 32 & 96   & 550 & 550 & 7.3\\
      H402 & 3.46 & 0.077 & 32 & 96   & 450 & 450 & 5.7\\
      \midrule                                                   
      N202 & 3.55 & 0.065 & 48 & 128 & 420 & 420 & 6.5\\
      N203 & 3.55 & 0.065 & 48 & 128 & 340 & 440 & 5.4\\
      N200 & 3.55 & 0.065 & 48 & 128 & 280 & 460 & 4.4\\
      D200 & 3.55 & 0.065 & 64 & 128 & 200 & 480 & 4.2\\
      \midrule
      N300 & 3.70 & 0.050 & 48 & 128 & 420 & 420 & 5.1\\
      J303 & 3.70 & 0.050 & 64 & 192 & 260 & 260 & 4.1\\
      \bottomrule
    \end{tabular}
    \caption{\label{tab_ens} List of CLS $N_\mathrm{f}=2+1$ ensembles used in the present study. The second column corresponds to the inverse bare coupling, $\beta=6/g^2_0$. In the third and fourth columns, $N_\mathrm{s}$ and $N_\mathrm{t}$, refer to the spatial and temporal extent of the lattice. Approximate values of the pion and Kaon masses are provided~\cite{Bruno:2014jqa,Bruno:2016plf}.}
  \end{center}
\end{table}

We employ a mixed action approach where the sea sector consists of
the just described CLS setup while Wilson twisted mass fermions at
maximal twist are used in the valence sector~\cite{Herdoiza:2017}. This setup is free
from leading lattice artefacts proportional to the valence quark
masses and is therefore particularly useful for computations in the
charm sector~\cite{Bussone:2018wki,Bussone:2018ljj}.
In Section~\ref{matching}, we provide a description of the valence quark action
and of the two strategies used to match the sea and valence quark masses.
Numerical results for continuum-limit scaling in this setup are shown in Section~\ref{cont}, while in Section~\ref{mixed_action} we provide additional evidence for the control of lattice artefacts and unitarity violations in this mixed action approach.

\section{Matching conditions}	\label{matching}

A chirally rotated mass term~\cite{Frezzotti:2000nk,Frezzotti:2004wz, Pena:2004gb, Sint:2007ug, Shindler:2007vp}, $\bm{ \mu_0 }=\text{diag} \left( \mu_{0,\mathrm{\ell}}, \mu_{0,\mathrm{\ell}}, \mu_{0,\mathrm{s}}, \mu_{0,\mathrm{c}} \right)$, is added to the Wilson operator in the valence sector as follows:
\begin{equation}
\frac{1}{2}\sum_{\mu=0}^3 \{\gamma_\mu(\nabla^*_\mu+\nabla_\mu) -a\nabla^*_\mu\nabla_\mu\} + \frac{i}{4} a c_{\mathrm{SW}} \sum_{\mu,\nu=0}^3 \sigma_{\mu\nu}\widehat F_{\mu\nu}+ \bm{m}_0 + i \gamma_5\, \bm{ \mu_0 }\,.
\end{equation}

Maximal twist is achieved by tuning on each ensemble the valence PCAC light quark mass, $\left. m_{12}^{\mathrm{R}} \right|_\mathrm{v}$, to zero by a linear interpolation using a set of values of the valence hopping parameter $ \left. \kappa_{\ell} \right|_\mathrm{v}$.\,\footnote{In what follows, the notation ``$|_\mathrm{v}$'' denotes the valence sector while ``$|_\mathrm{s}$'' refers to the sea sector. For quark masses, the subscripts $1$ and $2$ refer to two distinct light-quark flavours with degenerate masses, $m_1=m_2$, and the presence of a superscript ${\rm R}$ denotes renormalised quantities.}

\begin{figure}
  \centering
  \includegraphics[scale=0.4]{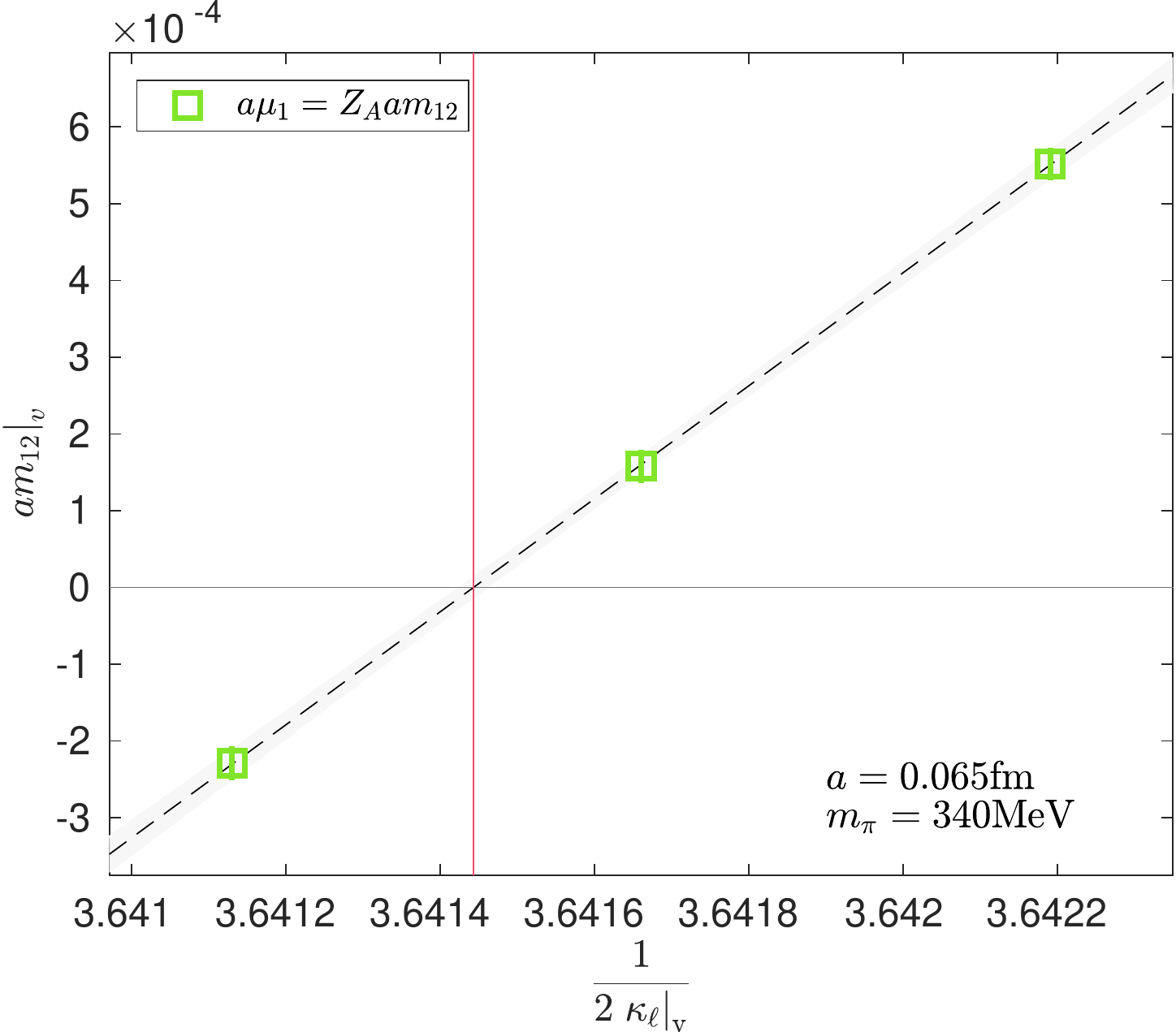}
  \caption{Illustration of the tuning to maximal twist by matching $\left. \mu_{1}^\mathrm{R} \right|_\mathrm{v} \; \equiv \; \left. m_{12}^\mathrm{R} \right|_\mathrm{s}$ in ensemble N203.}
  \label{plot:match_mq}
\end{figure}

\begin{figure}
	\centering
    \begin{subfigure}[b]{0.4\textwidth}
        \includegraphics[width=\textwidth]{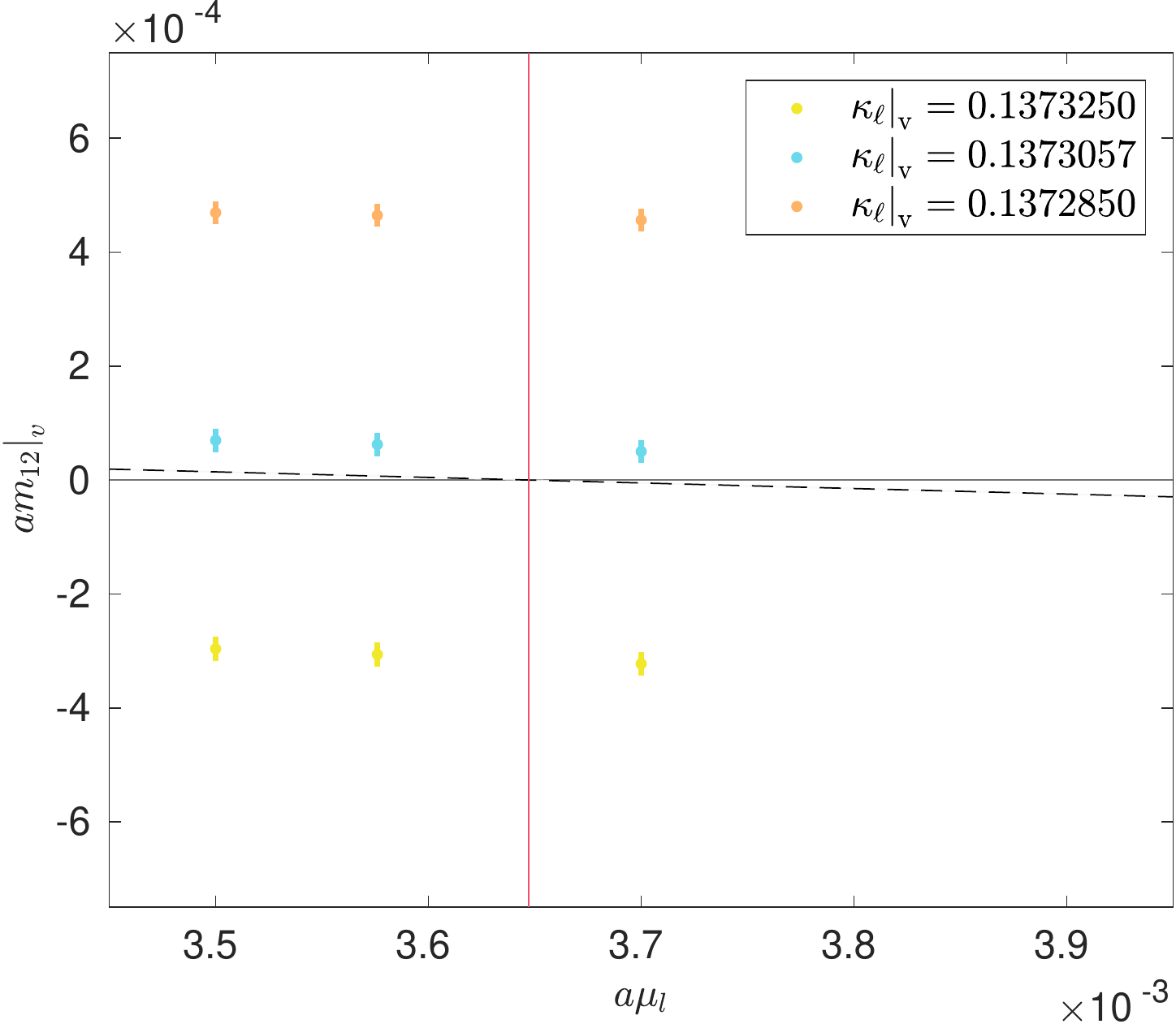}
        \caption{Valence PCAC quark mass as a function of the twisted mass in the neighbourhood of maximal twist for different values of $\kappa_{\ell}\rvert_\mathrm{v}$.}
        \label{plot:match_mps_mpcac}
    \end{subfigure}%
	\qquad \qquad
    \begin{subfigure}[b]{0.4\textwidth}
        \includegraphics[width=\textwidth]{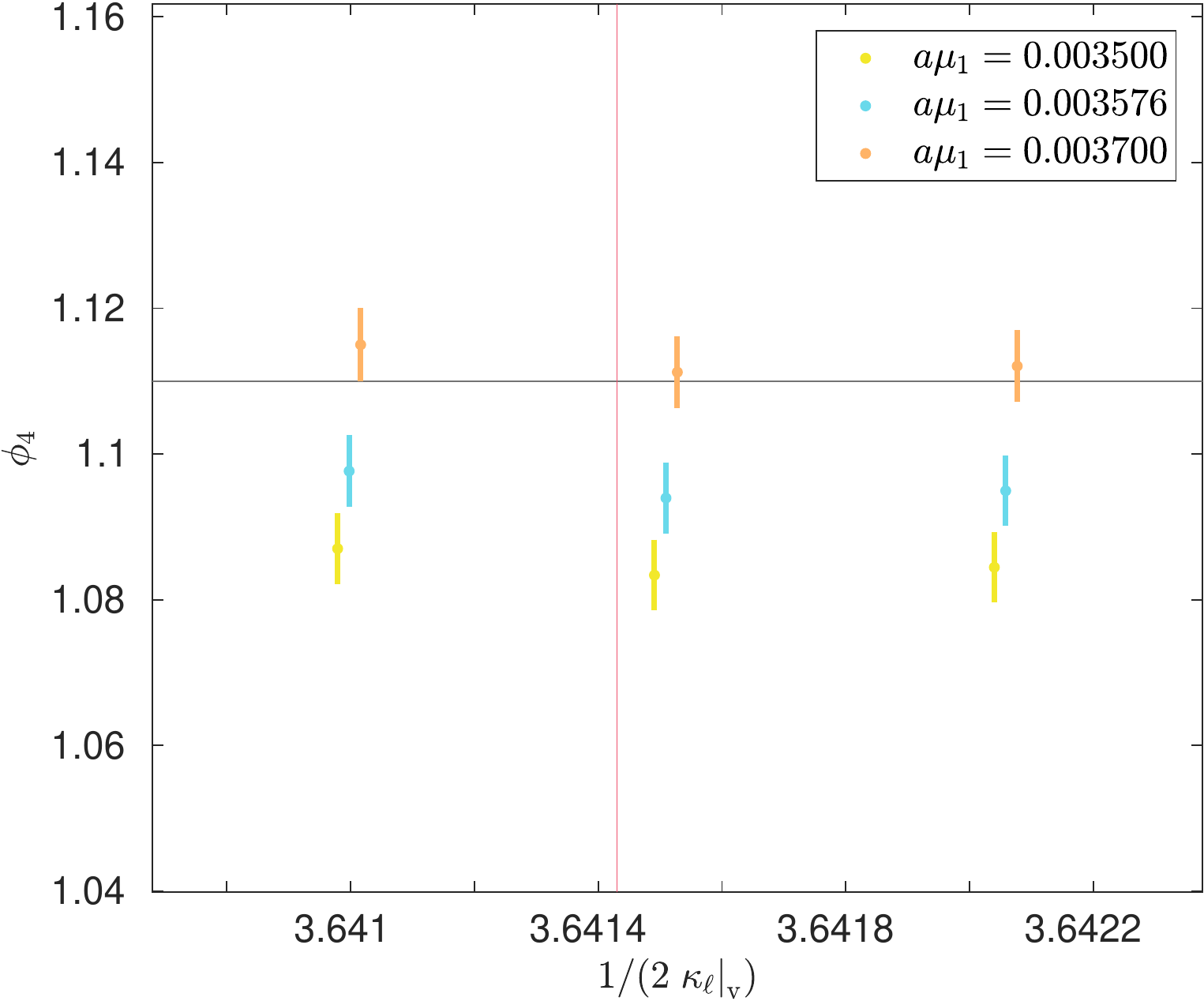}
        \caption{$\phi_4$ as a function of $ \frac{1}{2\kappa_{\ell}\rvert_\mathrm{v}}$ for different values of the twisted mass. 
        The horizontal line indicates the matching point.}
        \label{plot:match_mps_mps}
    \end{subfigure}%
    \caption{Illustration of the tuning to maximal twist by matching $\left. m_\pi \right|_\mathrm{v} \; \equiv \; \left. m_\pi \right|_\mathrm{s}$ for the ensemble N203.}
    \label{plot:match_mps}
\end{figure}

Furthermore, in order to recover unitarity in the continuum limit, it is necessary to match the sea and valence quark masses. In this work, two different procedures are analyzed: (i) matching of the renormalized $(u,d)$ and $s$ quark masses and (ii) matching of the pion and kaon masses.
The first method~\cite{Herdoiza:2017} is based on the matching of the renormalised PCAC quark mass in the sea sector to the renormalised twisted mass, $\left. \mu_{1}^\mathrm{R} \right|_\mathrm{v} \; \equiv \; \left. m_{12}^\mathrm{R} \right|_\mathrm{s}$. Including ${\rm O}(a)$ counterterms in the sea sector, this matching reads,
\begin{equation}
  \label{eq:match_mq}
  \frac{1}{Z_\mathrm{P}}\, \mu_1 \equiv
  \frac{Z_\mathrm{A}}{Z_\mathrm{P}}\, \left. m_{12} \right|_\mathrm{s}
  \left( 1 + \left( \tilde{b}_\mathrm{A} - \tilde{b}_\mathrm{P}
  \right) a \left. m_{12} \right|_\mathrm{s} + \left(
  \overline{b}_\mathrm{A} - \overline{b}_\mathrm{P} \right) a\,
  \mathrm{tr} \left. M_\mathrm{q} \right|_\mathrm{s} \right)\, ,
\end{equation}
where the renormalisation factor $Z_\mathrm{P}$ can be ignored since it appears on both sides of the equation.
This method allows to tune to maximal twist with just a few simulations around $\left. m_{12}^{\mathrm{R}} \right|_\mathrm{v} = 0$ as shown in Fig.~\ref{plot:match_mq}.
However, this approach relies on a determination of the mass-dependent
$b$-type improvement coefficients.\,\footnote{
We employ the non-perturbative determination in Ref.~\cite{Fritzsch:ongoing}. Further perturbative and non-perturbative studies have also appeared in Ref.~\cite{Taniguchi:1998pf} and Refs.~\cite{Korcyl:2016ugy,deDivitiis:2017vvw,Fritzsch:2018zym}, respectively.}
We employ the recent determination of $Z_A$ based on the chirally rotated Schr\"odinger functional~\cite{DallaBrida:2017}. 
A similar matching condition as in eq.~(\ref{eq:match_mq}) is used for the strange
quark mass.
An alternative way to match the light quark masses is given by the condition $\left. m_\pi \right|_\mathrm{v} \; \equiv \; \left. m_\pi \right|_\mathrm{s}$. This approach involves simulations on a {\it grid} of points in the $\left(\kappa_{\ell}\rvert_\mathrm{v}, \mu_{0,\mathrm{\ell}}\right)$ plane. 
The desired values of $\kappa_{\ell}\rvert_\mathrm{v}$ and $\mu_{0,\mathrm{\ell}}$ fulfilling, simultaneously, the matching and the maximal twist conditions are
obtained through interpolations over the {\it grid} of points.
The \textit{grid} can be efficiently chosen if the values of the twisted and the standard masses are selected based on the previous matching method. 
In such a small range of parameter space for $\left(\kappa_{\ell}\rvert_\mathrm{v}, \mu_{0,\mathrm{\ell}}\right)$, the valence PCAC quark mass and the pseudoscalar mass squared can be parametrized through low order polynomials in $\kappa_{\ell}\rvert_\mathrm{v}^{-1}$ and $\mu_{0,\mathrm{\ell}}$:\,\footnote{We notice that, in practice, $ \phi_4 $ and $ \phi_2 $ in eq.~(\ref{phi}), are used to perform the matching of the sea and valence quark masses.
}
\begin{align}
	\label{eq:par_light2}	
\left. m_{12} \right|_\mathrm{v} \left(\kappa_{\ell}\rvert_\mathrm{v}, \mu_{0,\mathrm{\ell}}\right) & =
		    \frac{p_{1,1}}{ 2\kappa_{\ell}\rvert_\mathrm{v} } 
		+ p_{1,2}  \mu_{0,\mathrm{\ell}}
		+ p_{1,3}
		\equiv 0, \\
  \label{eq:par_light}
\left. m_\pi^2 \right|_\mathrm{v} \left(\kappa_{\ell}\rvert_\mathrm{v}, \mu_{0,\mathrm{\ell}}\right) & =
		   \frac{p_{2,1}}{\left( 2\kappa_{\ell}\rvert_\mathrm{v} \right)^2 }  
		+  \frac{p_{2,2}}{2\kappa_{\ell}\rvert_\mathrm{v}} 
		+ p_{2,3} \mu_{0,\mathrm{\ell}}
		+ p_{2,4}
		\equiv \left. m_\pi^2 \right|_\mathrm{s}.	
\end{align}
Although a larger set of simulations are required with respect to the
matching of the quark masses in eq.~(\ref{eq:match_mq}), a benefit of this procedure is that only ${\rm O}(a)$-improved quantities appear in eqs.~(\ref{eq:par_light}) and~(\ref{eq:par_light2}).
Furthermore, the availability of a {\it grid} of points allows to incorporate a refined analysis of the mass-shifts towards a renormalised chiral trajectory.
An example of the application is shown in Fig.~\ref{plot:match_mps}.

The strange quark masses can be matched by imposing $\left. m_K \right|_\mathrm{v} \; \equiv \; \left. m_K \right|_\mathrm{s}$. 
In the neighborhood of the target strange quark mass, the valence kaon mass squared can be parametrized in the following way
\begin{align}
  \label{eq:par_kaon}
\left. m_K^2 \right|_\mathrm{v} \left(\kappa_{\ell}\rvert_\mathrm{v}, \mu_{0,\mathrm{\ell}}, \mu_{0,\mathrm{s}} \right) & =
		    \frac{p_{3,1}}{\left( 2\kappa_{\ell}\rvert_\mathrm{v} \right)^2 }
		+   \frac{p_{3,2}}{2\kappa_{\ell}\rvert_\mathrm{v}}
		+ p_{3,3}  \mu_{0,\mathrm{\ell}}
		+ p_{3,4}  \mu_{0,\mathrm{s}}
		+ p_{3,5}
		\equiv \left. m_K^2 \right|_\mathrm{s} .
\end{align}

\section{Continuum-limit scaling and light-quark mass dependence}	\label{cont}

\begin{figure*}
\centering
\begin{subfigure}[b]{0.475\textwidth}
    \centering
    \includegraphics[width=\textwidth]{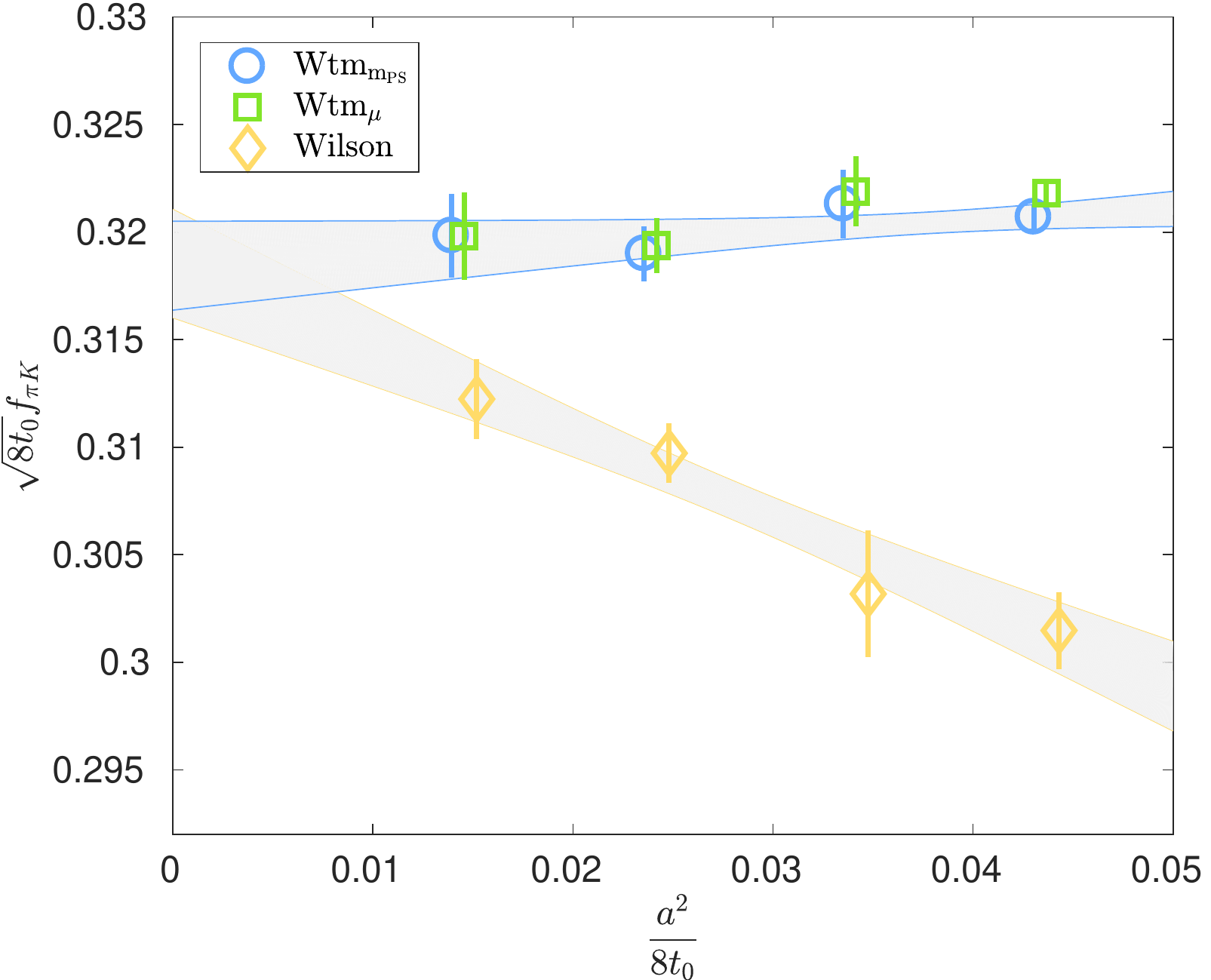}
	\caption[]%
	{{\small Continuum extrapolation of $f_{\pi K}$ for symmetric point ensembles.}}    
    \label{fig:fpiK_cont}
\end{subfigure}
\hfill
\begin{subfigure}[b]{0.475\textwidth}
    \centering 
    \includegraphics[width=\textwidth]{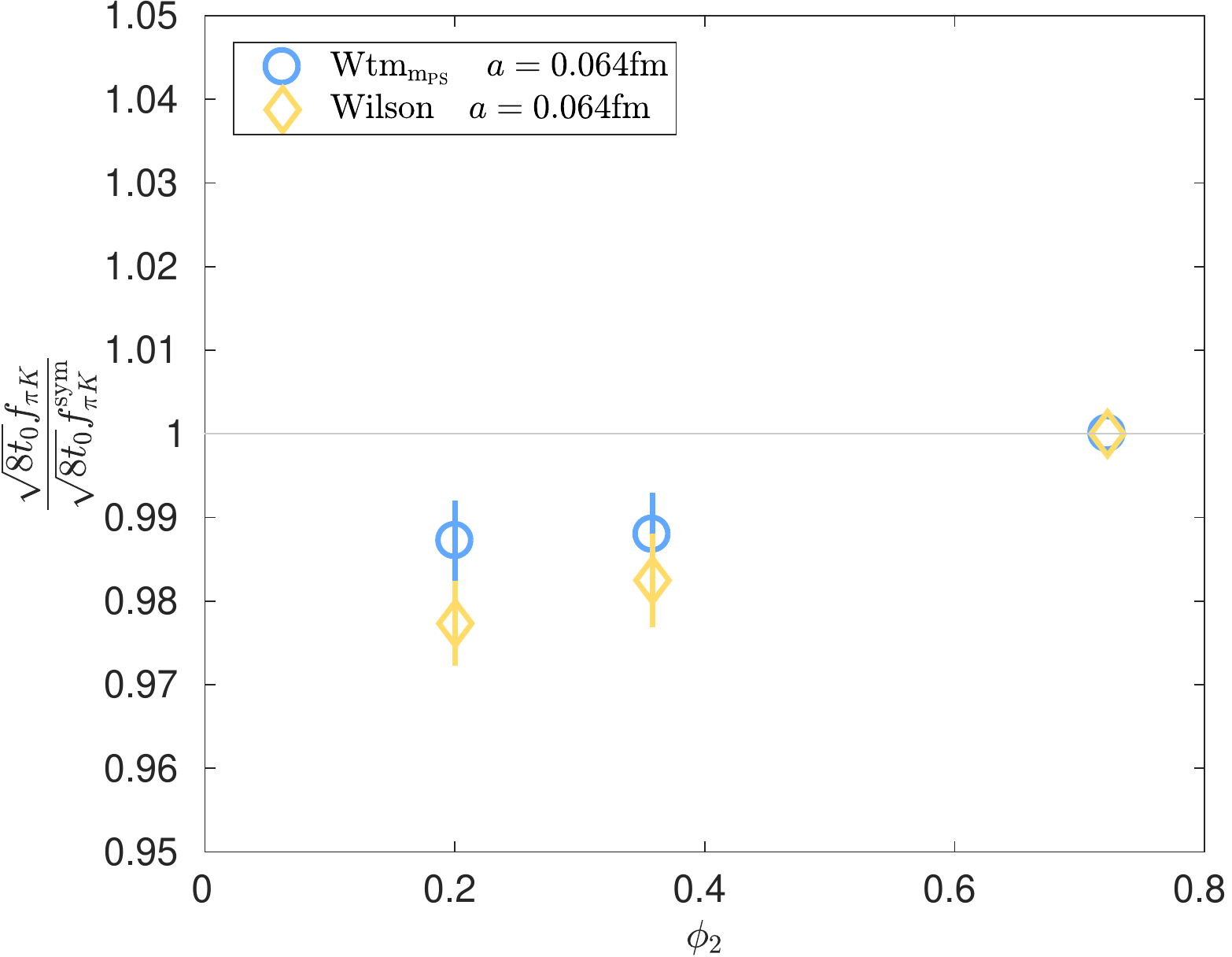}
    \caption[]%
    {{\small Light quark mass dependence of $f_{\pi K}$ at $\beta = 3.55$.}}
    \label{fig:mfpiK/fpiK_sym}
\end{subfigure}
\vskip\baselineskip
\begin{subfigure}[b]{0.475\textwidth}   
    \centering 
    \includegraphics[width=\textwidth]{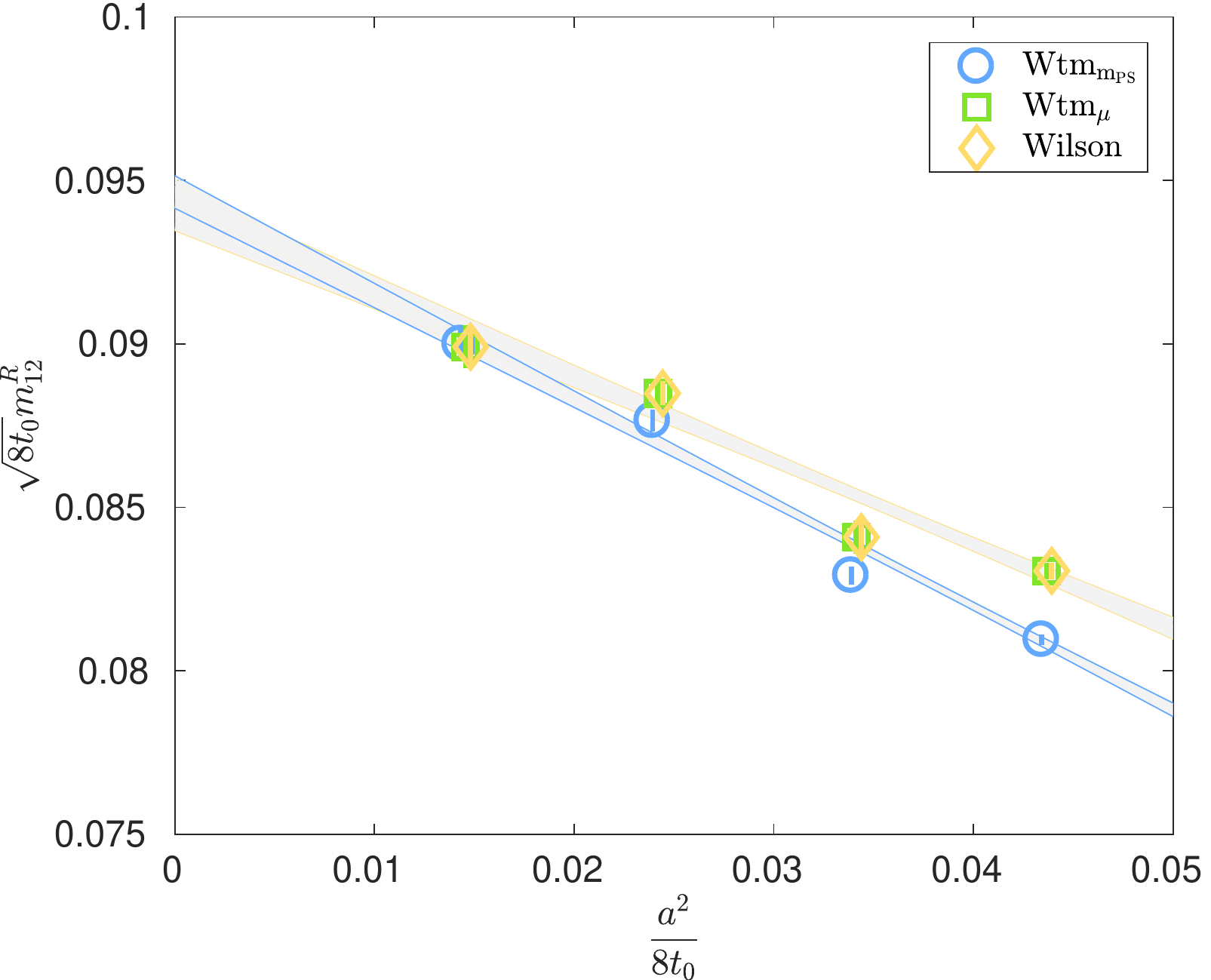}
    \caption[]%
    {{\small Continuum extrapolation of the light quark mass for symmetric point ensembles.}}    
    \label{fig:mq_cont}
\end{subfigure}
\quad
\begin{subfigure}[b]{0.475\textwidth}
    \centering 
    \includegraphics[width=\textwidth]{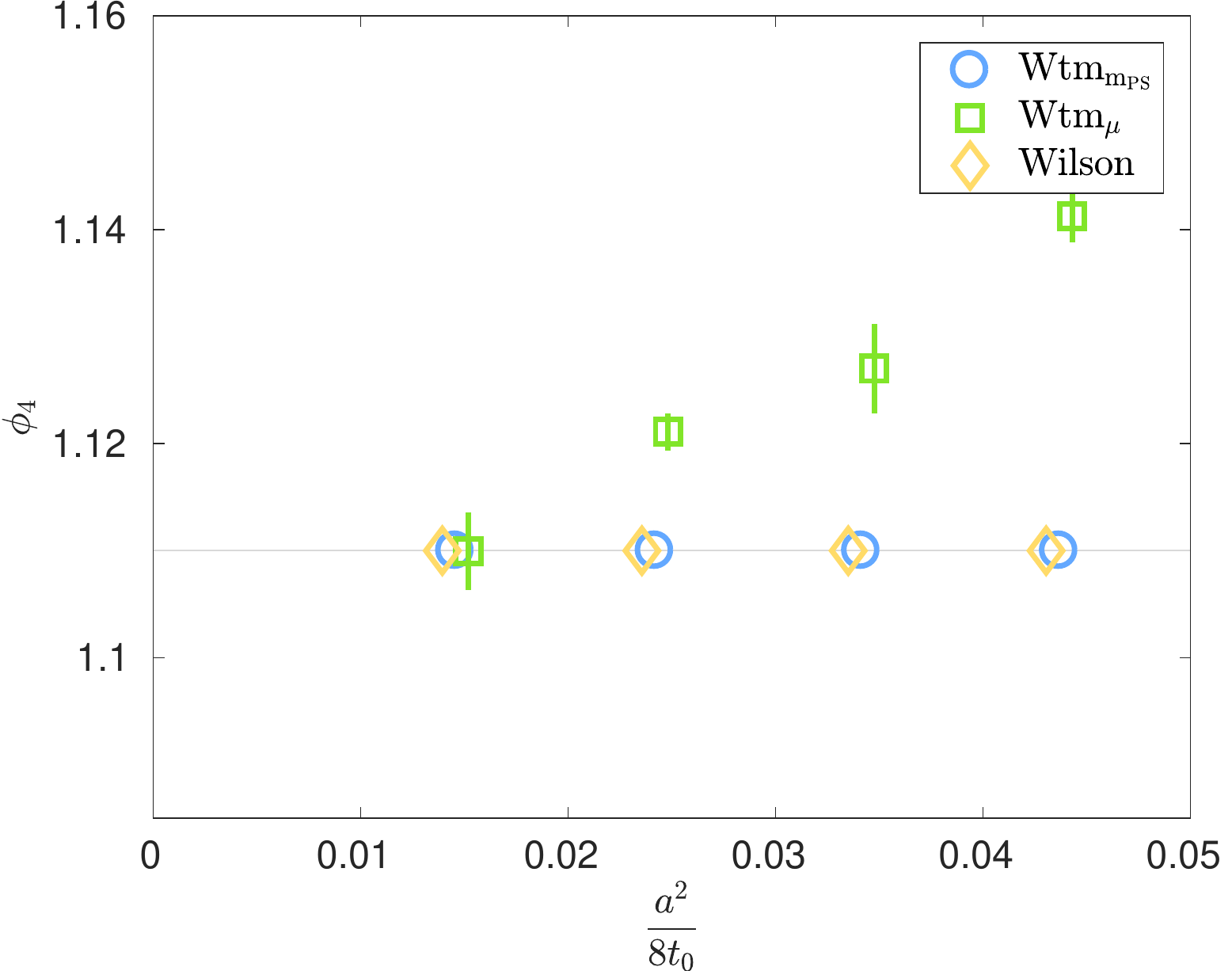}
    \caption[]%
    {{\small Lattice spacing dependence of $\phi_4$ for symmetric point ensembles.}}
    \label{fig:phi4_match}
\end{subfigure}
\caption{Comparison of results based on the Wilson and on the twisted mass
regularisatios using two matching procedures (${\rm Wtm}_{{\rm m}_{\rm PS}}$ and ${\rm Wtm}_{\mu}$).}
{\small }
\label{fig:cont_ext}
\end{figure*}

In Fig.~\ref{fig:fpiK_cont}, we illustrate the continuum-limit scaling of $f_{\pi K} = \frac{2}{3} \left( \frac{1}{2} f_{\pi} + f_K \right)$ in units $t_0$, where the pion and kaon decay constants $f_\pi$ and $f_K$, respectively, were computed as described in~\cite{Herdoiza:2017}. Three different sets of points are shown. In the legend, "Wilson" points refer to different ensembles at the symmetric point computed with the Wilson regularisation, whereas "Wtm" signals the use of Wilson twisted mass fermions at maximal twist. ${\rm Wtm}_{{\rm m}_{\rm PS}}$ and ${\rm Wtm}_{\mu}$ refer to the matching with the pseudoscalar mass and the quark mass, respectively. Notice that the different datasets are slightly shifted in the horizontal axis so they can be visualized.
This quantity exhibits a mild dependence on the matching condition and, furthermore, shows the agreement in the continuum limit among the two regularisations.

The quantity $f_{\pi K}$, normalized by its value at the symmetric point, shows a quadratic dependence on $\phi_2$, up to logarithmic corrections, as discussed in~\cite{Bruno:2016plf}.
The light quark mass dependence of $ \frac{f_{\pi K}}{ f_{\pi K}^{\rm sym}} $ computed with the Wtm regularisation by employing the matching with pseudoscalar meson masses, is shown in Fig.~\ref{fig:mfpiK/fpiK_sym} for three ensembles at a fixed value of the lattice spacing, $a=0.064$\,fm.

Figures~\ref{fig:mq_cont} and~\ref{fig:phi4_match} show the continuum-limit scaling of the RGI quark mass and of $\phi_4$, respectively. A comparison of the results for the Wilson and twisted regularisation using the two different matching procedures explained above are displayed.
The renormalisation and running of the quark mass in both regularisations are  based on Ref.~\cite{Campos:2018}.
Our results indicate that the difference between both matching procedures decreases for finer values of the lattice spacing.

\section{Additional checks of the mixed action}	\label{mixed_action}

\begin{figure}
  \centering
  \includegraphics[scale=0.73]{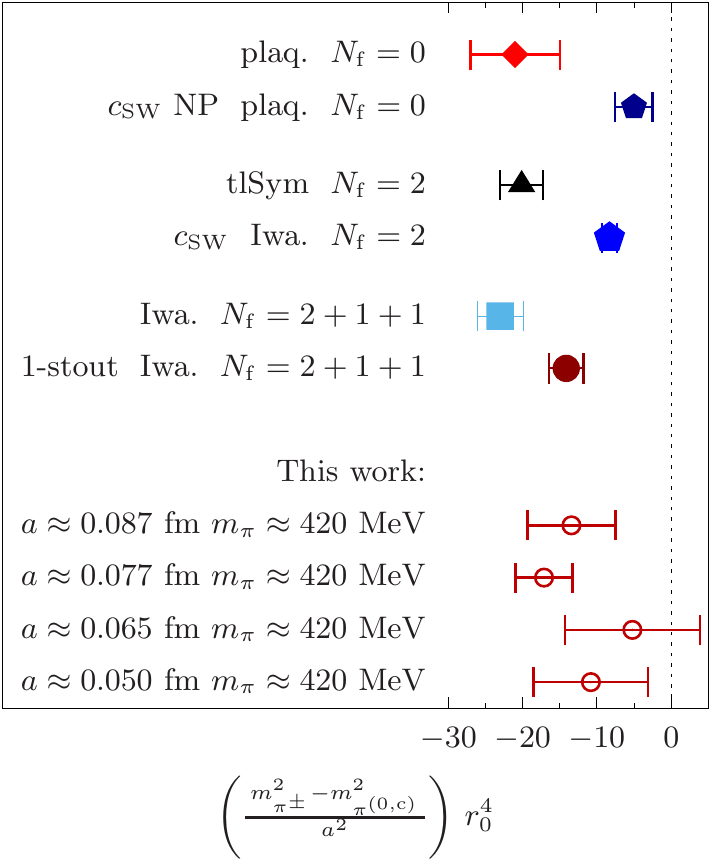}
  \caption{Comparison of the mass splitting between the charged and the neutral connected pions for various lattice actions~\cite{Herdoiza:2013sla, Abdel-Rehim:2015pwa}. Notice that only a small subset of the available configurations were used in this preliminary study	.}
 \label{plot:pion_mass_splitting}
\end{figure}

The mass splitting between the charged and neutral connected pions in units of the Sommer parameter $r_0$ measures the isospin breaking effects induced by the valence regularisation.
Although both masses coincide in the sea, the twisted mass valence action introduces a splitting proportional to the scale of $O(a^2)$ lattice artefacts. In Fig.~\ref{plot:pion_mass_splitting}, we compare our measurements of pion mass splitting at different values of the lattice spacing to various determination with different lattice actions~\cite{Herdoiza:2013sla, Abdel-Rehim:2015pwa}.
This Figure indicates that the isospin breaking effects in our mixed action setup are on the same ballpark as those from other lattice regularisations.

\section{Conclusions}	\label{conclusions}

We have described two matching conditions of sea and valence quark
masses of a mixed action, based on renormalized quark masses and on
pseudoscalar meson masses. Preliminary results for the continuum-limit
scaling for $f_{\pi K}$ and $m_{12}^R$ in terms of $t_0$ are presented together with additional consistency checks of the mixed action

The tuning procedure for the pseudoscalar mass matching requires simulating a \textit{grid} of points in the plane of the bare standard mass and the twisted mass.
In practice, this method also allows to shift the valence masses to a family of renormalised chiral trajectories of sea quark masses.

The pseudoscalar mass matching procedure is beneficial since it relies on quantities which are free from ${\rm O}(a)$ effects. Our preliminary results show that the overall difference between our matching procedures becomes smaller as the lattice spacing is reduced.

\subsection*{Acknowledgements}

We thank our CLS colleagues for producing the gauge configuration ensembles used in this study. 
We acknowledge PRACE for awarding us access to MareNostrum at the Barcelona
Supercomputing Center (BSC), Spain. We thank CESGA for granting access to FinisTerrae II.
We thankfully acknowledge support through the Spanish MINECO project FPA2015-68541-P, the Centro de Excelencia Severo Ochoa Programme SEV-2016-0597 and the Ram\'{o}n y Cajal Programme RYC-2012-10819.

\end{document}